# Structure and evolution of urban heavy truck mobility networks

Yitao Yang[1,3], Bin Jia[1,2,*], Erjian Liu[1,4], Xiao-Yong Yan[2,*], Michiel de Bok[3], Lóránt A. Tavasszy[3], Ziyou Gao[1,2,†]

[1] *Key Laboratory of Integrated Transport Big Data Application Technology for Transport Industry, Beijing Jiaotong University, Beijing 100044, China.*

[2] *Institute of Transportation System Science and Engineering, Beijing Jiaotong University, Beijing 100044, China.*

[3] *Department of Transport & Planning, Faculty of Civil Engineering and Geosciences, Delft University of Technology, Stevinweg1, Delft 2628 CN, the Netherlands*

[4] *Instituto de Física Interdisciplinar y Sistemas Complejos IFISC (CSIC-UIB), Palma de Mallorca 07122, Spain.*

Corresponding authors: [*] B. Jia (bjia@bjtu.edu.cn), [*] X. Y. Yan (yanxy@bjtu.edu.cn), [†] Z. Gao (zygao@bjtu.edu.cn)

Revealing the structural properties and understanding the evolutionary mechanisms of the urban heavy truck mobility network (UHTMN) provide insights in assessment of freight policies to manage and regulate the urban freight system, and are of vital importance for improving the livability and sustainability of cities. Although massive urban heavy truck mobility data become available in recent years, in-depth studies on the structure and evolution of UHTMN are still lacking. Here we use massive urban heavy truck GPS data in China to construct the UHTMN and reveal its a wide range of structure properties. We further develop an evolving network model that simultaneously considers weight, space and system element duplication. Our model reproduces the observed structure properties of UHTMN and helps us understand its underlying evolutionary mechanisms. Our model also provides new perspectives for modeling the evolution of many other real-world networks, such as protein interaction networks, citation networks and air transportation networks.

**Introduction**

Modern city is the center of commodity production and consumption, and is supported by massive goods movements. Urban freight vehicle mobility between functional locations, e.g., industrial companies, logistics warehouses and comprehensive markets, establishes spatial interactions between different urban regions[1,2]. Heavy trucks are the most frequently observed type of urban freight vehicles, undertaking high-volume transportation tasks between urban functional locations and playing an irreplaceable role in the urban economy[3]. For example, although heavy trucks account for less than 40% of urban freight vehicles, they carry more than 80% of the freight volume in Melbourne[4]; the U.S. heavy truck related transportation sectors generate more than 500 billion dollars in annual tax revenue and provides jobs for 10 million people[5]. Meanwhile, heavy truck mobility brings non-negligible negative impacts, such as traffic congestion, accidents and air pollution, raising a need for evidence-based management and regulation of urban freight systems to promote urban sustainable development[6]. Studying the global properties and dynamical mechanisms of urban heavy truck mobility helps us understand the architecture and evolution of the urban freight systems, which are of vital importance for the design of measures to manage and regulate urban freight systems[7]. Until recently a lack of real data had limited our in-depth understanding of urban heavy truck mobility at the network level. However, the advent of the big data era provides the possibility to explore the global properties and dynamical mechanisms of urban heavy truck mobility from the perspective of complex networks[8].

The use of complex networks theory to study urban mobility has become widespread[9-12], where the structure and evolution of mobility networks are two fundamental research issues. Quantitatively characterizing the structure of mobility networks helps us understand the global properties of urban mobility, and exploring the evolutionary process of mobility networks enables us to reveal the dynamical mechanisms underlying the global properties of urban mobility. In recent years, a great deal of effort has been devoted to the study of urban human mobility[8,13], providing the theoretical background for studying the structure and evolution of urban heavy truck mobility network (UHTMN). For quantitatively characterizing the structure of mobility networks, most previous studies have used massive human mobility data, such as mobile phone data[14], GPS trajectory data[15], social network data[16], and smart card data[17] to obtain origin-destination (OD) matrices and construct human mobility networks at different scales[18,21]. Previous studies characterized a wide range of network structure properties, including the scale-free[22], small-world[23], network similarity[24] and network hierarchy[25], within the theoretical framework of network science. In terms of freight mobility networks, previous studies focused on the large-scale inter-country or inter-city freight mobility networks[18,26-31], but the in-depth studies on urban freight mobility networks are still scarce. In the era of big data, massive urban heavy truck mobility data becomes available[32], providing us the possibility to construct UHTMN and characterize its structural properties.

Regarding the study of the dynamical mechanisms of mobility networks, the research goal is to develop an evolving network model to reproduce the observed properties of mobility network structure and to explain the underlying dynamical mechanisms[33,34]. Early studies mainly concentrated on the unweighted evolving network models, and the most representative one is the Barabasi-Albert (BA) model[22]. The BA model propose a preference attachment rule and reproduce the scale-free property of real-world networks, providing a theoretical explanation for the richer-get-richer phenomenon[35]. In addition, there are many other unweighted evolving network models[36,37] that have subsequently been developed. However, these models only consider network topology growth but ignore the network edge weight. In UHTMN, edge weight indicates the interaction strength between locations, which is a key feature of UHTMN. Therefore, it is essential to consider edge weight in modeling the evolution of UHTMN. Previous studies developed many weighted evolving network models[38-41] from different perspectives, and the most representative one is the Barrat-Barthelemy-Vespignani (BBV) model[41]. The BBV model incorporates a strength preference attachment rule and a weight update rule to reproduce the traffic increment phenomenon. This model has a wide range of applications, but ignore another important feature, i.e., space. The nodes of many real-world networks, including the UHTMN, are located in geographic space, implying that the interactions of node pairs are constrained by the cost associated with space[8]. To explain the evolutionary mechanisms of spatial networks, previous studies[42-45] proposed many evolving network models based on the network growth rule of preference attachment with spatial constraint. Taken together, the above three classes of evolving network models all follow the preference attachment rule in network growth, but this rule cannot explain the phenomenon of system element duplication, which is widely observed in many real-world networks[46-48]. For example, the findings of new research are usually built on previous pioneering studies, and this is considered to be the key mechanism underlying the evolution of citation networks[47]; the function of new proteins is usually considered to derive from the duplication of original genes and follows the structural divergences, which determine the dynamic evolution of protein interaction network[48]. Similar to these real-world systems, the establishment of a new company, i.e., a node of UHTMN, is usually accompanied by duplicating the business modes of existing companies[49,50], which is a potential factor that drives the evolution of UHTMN. Therefore, we need to develop an evolving network model that considers not only the weight and space, but also the system element duplication.

The goal of this paper is to quantitatively characterize the structural properties of UHTMN and develop an evolving network model to explain the evolutionary mechanisms of UHTMN. We use urban heavy truck mobility data and freight-related point-of-interest (POI) data to construct UHTMNs of typical cities in China. For quantitatively characterizing the UHTMN structure, we first analyze the global topology properties and node neighbor similarity property, and then analyze the heterogeneity and hierarchy of network weight distribution. Next, we develop an evolving network model that simultaneously considers weight, space and system element duplication to explain the evolutionary mechanisms of UHTMN. The model results show that our model can reproduce a wide range of structure properties of UHTMN. Finally, we analyze the effects of model parameters on the network structure properties and discuss the potential applications of our model in other real-world systems.

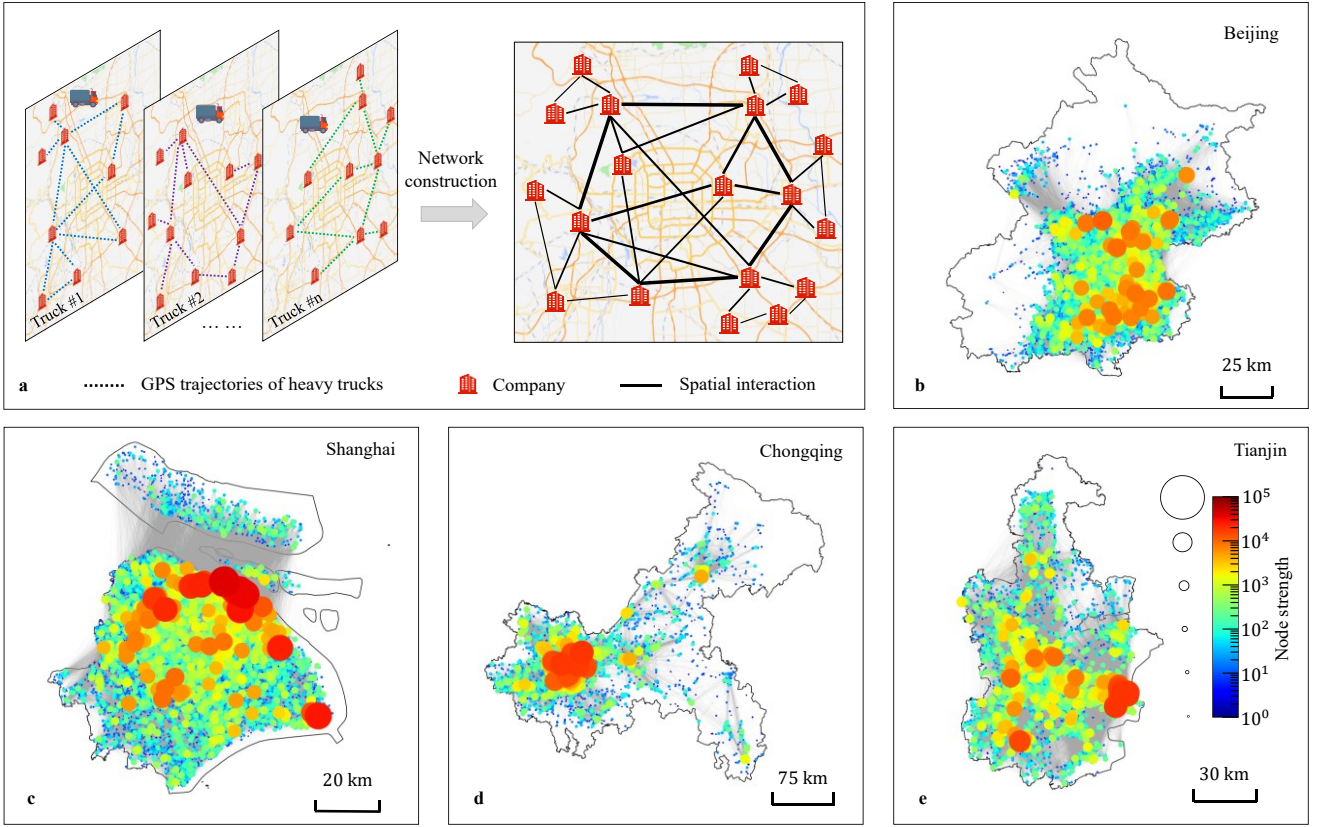

**Fig. 1.** Construction and illustration of urban heavy truck mobility networks. **a** Network construction by using heavy truck GPS trajectory data and urban freight-related POI data. Individual mobility network of each heavy truck (left panels) is first constructed by identifying the trip origins and destinations from GPS trajectories, and then the urban heavy truck mobility network (right panel) is constructed from heavy truck collective movements. Network nodes and edges are represented by urban freight locations and spatial interactions between them. The width of each line in the right panel indicates truck flow between two freight locations. **b-e** Illustration of empirical urban heavy truck mobility networks of four cities (municipalities directly under the central government) in China. Size and color of each node indicate its strength.

**Table 1.** Global characterization of heavy truck mobility networks of four cities (municipalities directly under the central government) in China. Number of nodes $N$; number of edges $E$; average degree $\langle k \rangle$; diameter $\phi$; average shortest path $\langle \ell \rangle$; average clustering coefficient $\langle C \rangle$; average shortest path $\langle \ell_{rand} \rangle$ and average clustering coefficient $\langle C_{rand} \rangle$ of ER random network that has the same number of nodes and edges as the UHTMN of specific city.

| City | $N$ | $E$ | $\langle k \rangle$ | $\phi$ | $\langle \ell \rangle$ | $\langle \ell_{rand} \rangle$ | $\langle C \rangle$ | $\langle C_{rand} \rangle$ |
|---|---|---|---|---|---|---|---|---|
| Beijing | 15983 | 207264 | 25.9 | 9 | 3.63 | 3.3 | 0.0318 | 0.0016 |
| Shanghai | 16094 | 436172 | 54.2 | 7 | 3.07 | 2.78 | 0.0777 | 0.0034 |
| Chongqing | 10629 | 203203 | 38.2 | 12 | 3.76 | 2.89 | 0.0757 | 0.0036 |
| Tianjin | 10697 | 216724 | 40.5 | 7 | 3.12 | 2.85 | 0.0668 | 0.0038 |

## Results

### Structure of urban heavy truck mobility networks

We construct the urban heavy truck mobility network (UHTMN) by using massive heavy truck GPS trajectory data and urban freight-related point-of-interest (POI) data (see **Fig. 1a-e**). The UHTMN is a weighted undirected network, in which node represents urban freight location; edge $E(i, j)$ represents the spatial interaction between freight location $i$ and $j$; the weight $w_{ij}$ of edge $E(i, j)$ represents the interaction strength between freight location $i$ and $j$, i.e., heavy truck flow. The details of the network construction are shown in the Method section.

We quantitatively characterize the various topological and weighted structure properties of empirical UHTMN. Here, we select the UHTMNs of four municipalities directly under the central government in China, i. e., Beijing, Shanghai, Chongqing and Tianjin, for analysis. The analysis results for the UHTMNs of eight additional typical cities in China are shown in **Supplementary Information**. We first analyze the topological structure properties of UHTMN without considering network weights. We calculate the global topological measures of UHTMNs of four Chinese municipalities (see **Supplementary Note 1** for the details of the measures), as shown in **Table 1**. The network sizes of UHTMNs vary across cities, but the UHTMNs of these cities have similar structure properties, i.e., the average shortest path length $\langle \ell \rangle$ of UHTMN is close to that $\langle \ell_{rand} \rangle$ of the corresponding ER random graph[51], which has the same number of nodes and edges as the UHTMN of specific city. And the average clustering coefficient $\langle C \rangle$ of UHTMN is much larger than that $\langle C_{rand} \rangle$ of the corresponding ER random graph. These results suggest that UHTMN has a small-world property[23], implying that urban freight locations form multiple well-connected local clusters, and some freight locations bridge the interactions of these local clusters.

Next, we calculate the degree of each node and obtain the degree distributions $p(k)$ of UHTMNs (see **Fig. 2a-d**), which obey the cut-off power-law distribution, i.e., $p(k) \sim (k+\Delta k)^{-\gamma} e^{-k/k_x}$, where $\gamma$ is the power exponent and $k_x$ is the cutoff value. These results suggest that UHTMN has a scale-free property[22], indicating that the connections of a few freight locations are much higher than those of other freight location. We then analyze the edge distance distributions $P(d)$ of UHTMNs (see **Fig. 2e-h**) and find they decay exponentially, i.e., $P(d) \sim e^{-d/r_d}$, where $r_d$ is the typical scale, reflecting the spatial feature of UHTMN. The cost of establishing long-distance connections between freight locations is greater than short-distance connections, and this limits the number of long-distance edges in the UHTMN. Besides, we analyze the rich club coefficient $RC(k)$ (see **Fig. 2i-l**), which is an increasing function of $k$ and closed to 1 for sufficiently large $k$. The results indicate that UHTMN has a rich-club property[35], implying that the freight locations with few connections tend to distribute in different local clusters and are poorly connected to each other, while the freight locations with many connections tend to play a role like global hubs. Moreover, we use the Jaccard similarity coefficient[52] to measure the similarity of two nodes in UHTMN (see **Method** for details). For each node in the network, we search for another node that is most similar to this node corresponding to the largest Jaccard coefficient, and they comprise a similar node pair. The two nodes in a similar node pair tend to share partners for the citation networks[47]; or have similar system functions for the biological networks[48]. We calculate the Jaccard similarity coefficient $Jac$ of each similar node pair and obtain their mean value $Avg_{real}$ and distribution $p(Jac)$, as shown in **Fig. 2m-p**. We find the $Avg_{real}$ in UHTMN is significantly larger than that in most other real-world networks (see **Fig. S2** for details). The results suggest that UHTMN has a node neighbor similarity property, implying that many freight locations tend to transport cargo or provide products to the same clients. Besides, the $p(Jac)$ is a bimodal distribution, in which the data of $Jac$ close to 1 is one of the probability peaks, indicating that many freight locations with few connections in local region tend to connect with each other.

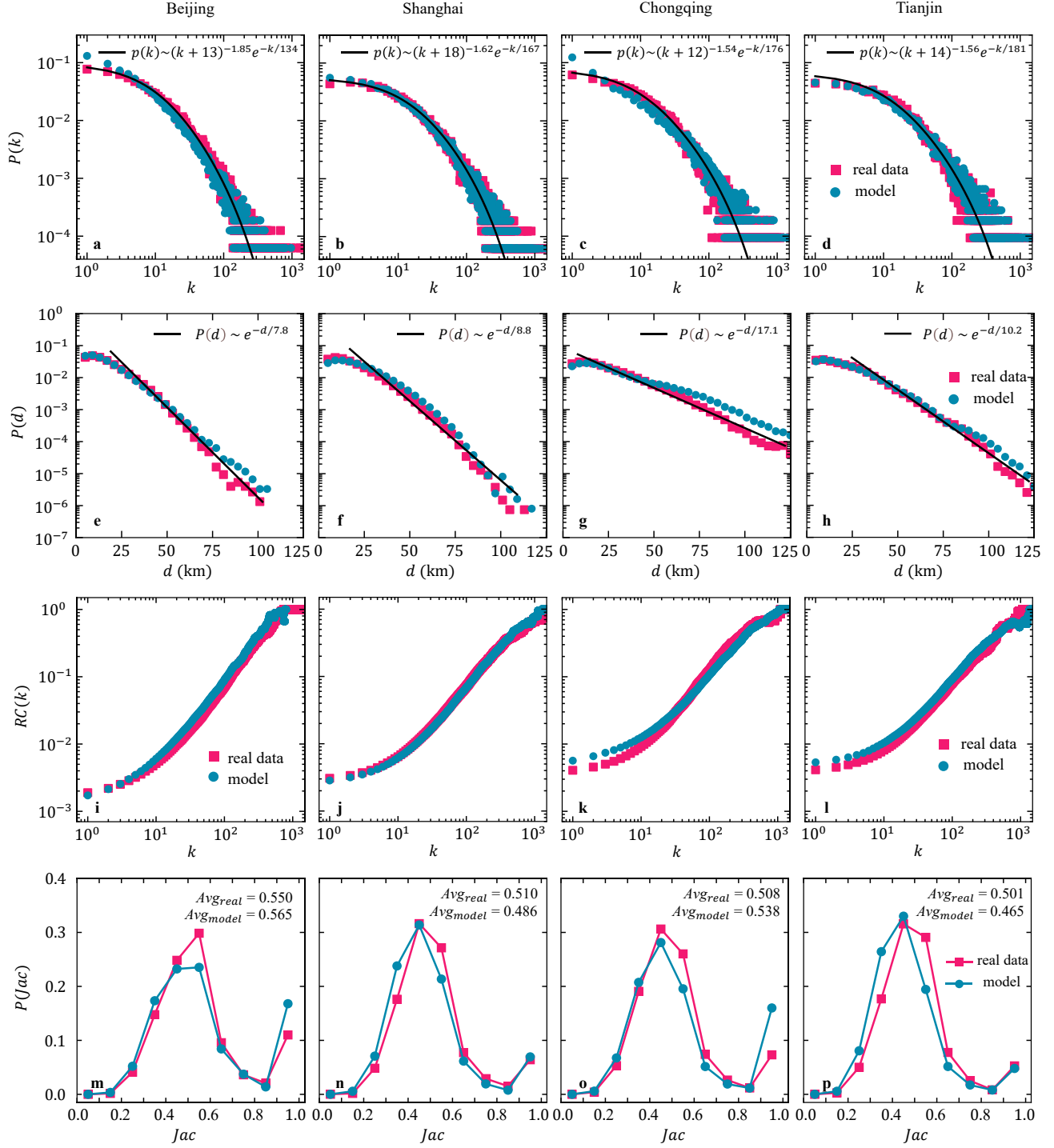

**Fig. 2.** Topological properties observed from four empirical UHTMNs and reproduced by the proposed evolving network model. **a-d** Degree distributions $p(k)$. The line represents the fitted cut-off power-law distribution $p(k) \sim (k + \Delta k)^{-\gamma} e^{-k/k_x}$. **e-h** Edge distance distributions $p(d)$. The line represents the fitted exponential distribution $P(d) \sim e^{-d/r_d}$. **i-l** Rich-club coefficient distributions $RC(k)$. **m-p** Distributions of Jaccard similarity coefficient of nodes and their similar nodes $p(Jac)$. $Avg_{real}$ is the mean value of Jaccard similarity coefficient for empirical UHTMN, and $Avg_{model}$ is the one for model network.

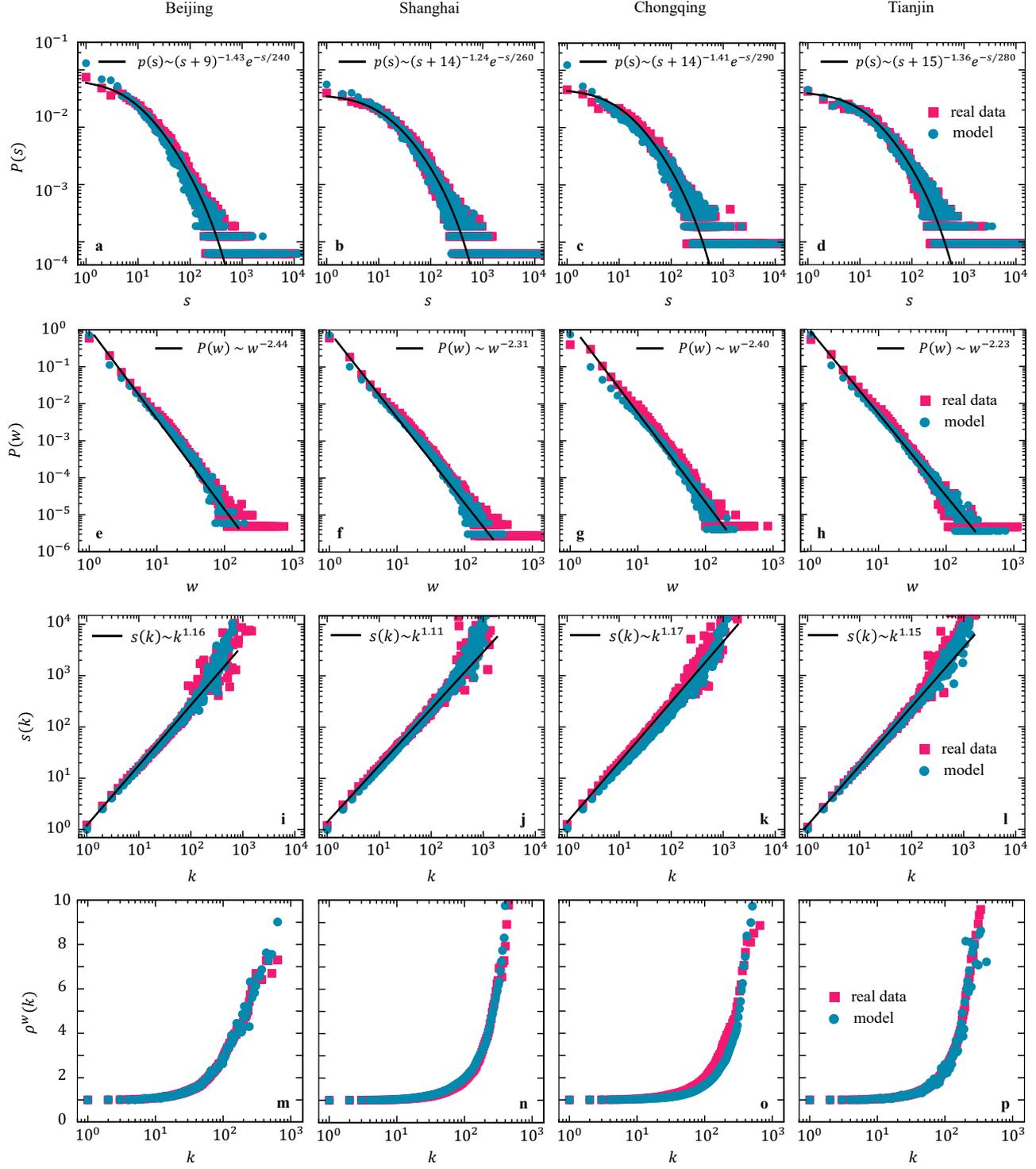

**Fig. 3.** Weight heterogeneity observed from four empirical UHTMNs and reproduced by the proposed evolving network model. **a-d** Node strength distributions $p(s)$. Each line represents the fitted cut-off power-law distribution, i.e., $p(s) \sim (s + \Delta s)^{-\gamma_s} e^{-s/s_x}$. **e-h** Edge weight distributions $p(w)$. Each line represents the fitted power-law distribution, i.e., $p(w) \sim w^{-\gamma_w}$. **i-l** Average strength $s(k)$ as a function of degree $k$. Each line represents the super-linear relation, i.e., $s(k) \sim k^\beta$ ($\beta > 1$). **m-p** Weighted rich-club coefficient distribution $\rho^w(k)$.

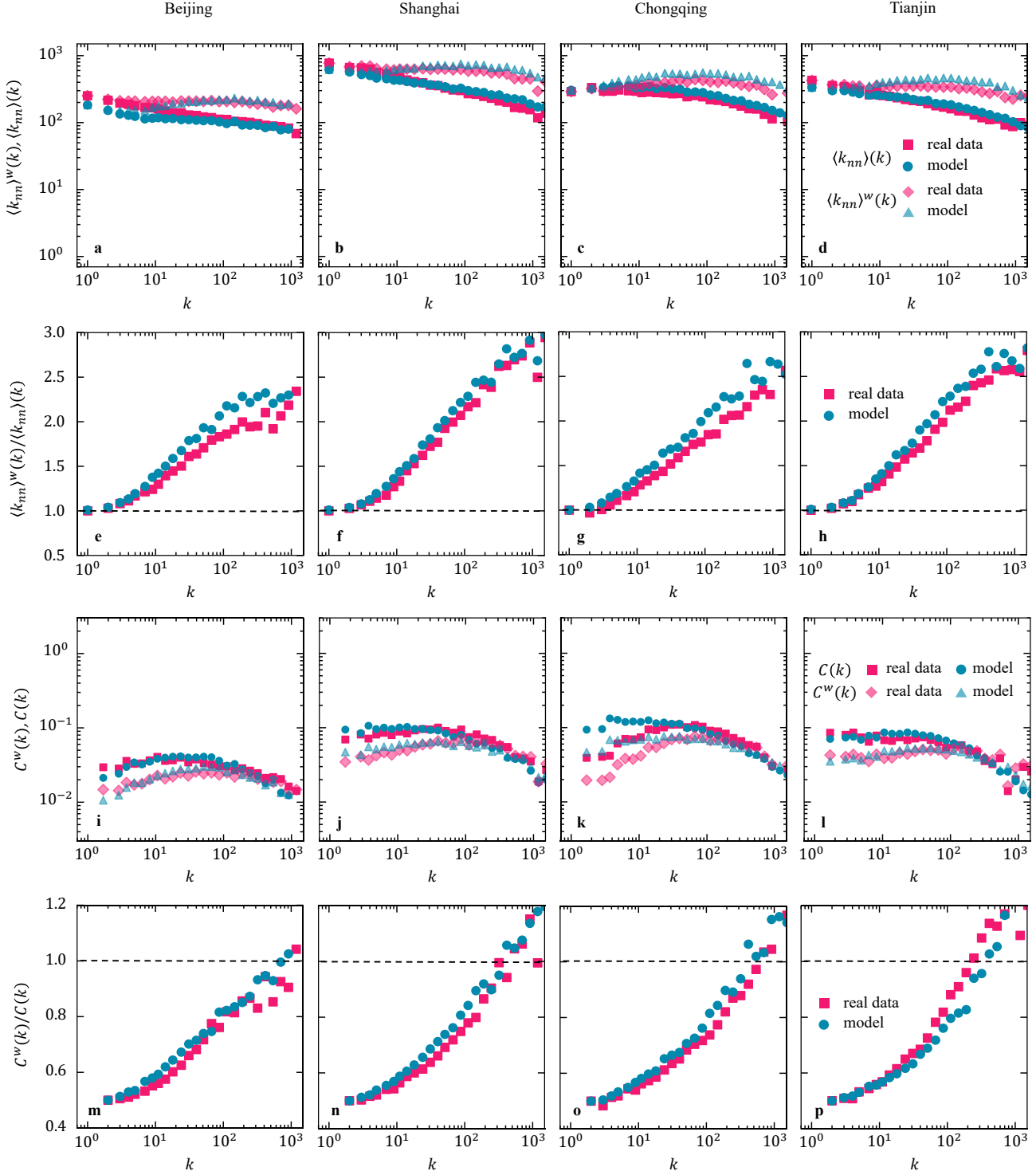

**Fig. 4.** Weight distributions among nodes with different degrees observed from four empirical UHTMNs and reproduced by the proposed evolving network model. **a-d** Distributions of weighted average degree of nearest neighbors $\langle k_{nn}\rangle^w(k)$ and unweighted one $\langle k_{nn}\rangle(k)$. **e-h** The ratio of $\langle k_{nn}\rangle^w(k)$ to $\langle k_{nn}\rangle(k)$, i.e., $\langle k_{nn}\rangle^w(k)/\langle k_{nn}\rangle(k)$. Each dashed line indicates $\langle k_{nn}\rangle^w(k)/\langle k_{nn}\rangle(k) = 1$. **i-l** Distributions of weighted clustering coefficient $C^w(k)$ and unweighted one $C(k)$. **m-p** The ratio of $C^w(k)$ to $C(k)$, i.e., $C^w(k)/C(k)$. Each dashed line indicates $C^w(k)/C(k) = 1$.

Furthermore, we characterize the weighted structure properties of UHTMN by analyzing the heterogeneous distribution of network weights. We first obtain the node strength distribution $p(s)$, which follows a power-law distribution, i.e., $p(s) \sim (s + \Delta s)^{-\gamma_s} e^{-s/s_x}$ (see **Fig. 3a-d**), where $\gamma_s$ is the power exponent, indicating that the truck flows into and out of some freight locations are significantly higher than those of other freight locations.

We then obtain the edge weight distribution $p(w)$, which also follows a power-law distribution, i.e., $p(w) \sim w^{-\gamma_w}$ (see **Fig. 3e-h**), where $\gamma_w$ is the power exponent, indicating that truck flows between a small number of freight locations are significantly high. Besides, we find that the mean value of node strength $s(k)$ and degree $k$ have a super-linear scaling relationship, i.e., $s(k) \sim k^\beta$ ($\beta > 1$) (see **Fig. 3i-l**), indicating that the truck flows between a freight location and other interacting freight locations tend to be positively correlated with the connections of this freight location. Additionally, the weighted rich-club coefficient $\rho^w(k)$ is an increasing function of degree $k$ (see **Fig. 3m-p**), indicating that freight locations with many connections are not only well connected (see **Fig. 2i-l**), but also have significantly high truck flows between them.

Finally, we analyze the heterogeneous distribution of network weights by using two measures, i.e., average degree of nearest neighbors and clustering coefficient. We first calculate the unweighted average degree of nearest neighbors $\langle k_{nn} \rangle(k)$ and the weighted one $\langle k_{nn} \rangle^w(k)$. We find that $\langle k_{nn} \rangle(k)$ is the decreasing function of $k$ (see **Fig. 4a-d**), indicating the disassortativity of UHTMN, i.e., freight locations with few connections tend to establish interactions with freight hub locations. Besides, $\langle k_{nn} \rangle^w(k) \geq \langle k_{nn} \rangle(k)$ at different values of $k$ and the ratio of $\langle k_{nn} \rangle^w(k)$ to $\langle k_{nn} \rangle(k)$, i.e., $\langle k_{nn} \rangle^w(k)/\langle k_{nn} \rangle(k)$, is an increasing function of $k$ (see **Fig. 4e-h**), suggesting that truck flows between two freight locations tend to be positively correlated with their connections, as shown in **Fig. S1b-e**. And then we calculate the weighted and unweighted clustering coefficient, i.e., $C^w(k)$ and $C(k)$. We find that both $C^w(k)$ and $C(k)$ decay in a power-law behavior at large $k$ (see **Fig. 4i-l**); and $C^w(k)/C(k)$ is the increasing function of $k$ and $C^w(k) \geq C(k)$ at large $k$, while $C^w(k) < C(k)$ at small $k$ (see **Fig. 4m-p**). The results suggest that freight locations with few connections tend to form multiple well-connected local clusters and the truck flows between them are commonly small; in other words, truck flows between two freight locations tend to be positively correlated with their connections, as shown in **Fig. S1b-e**.

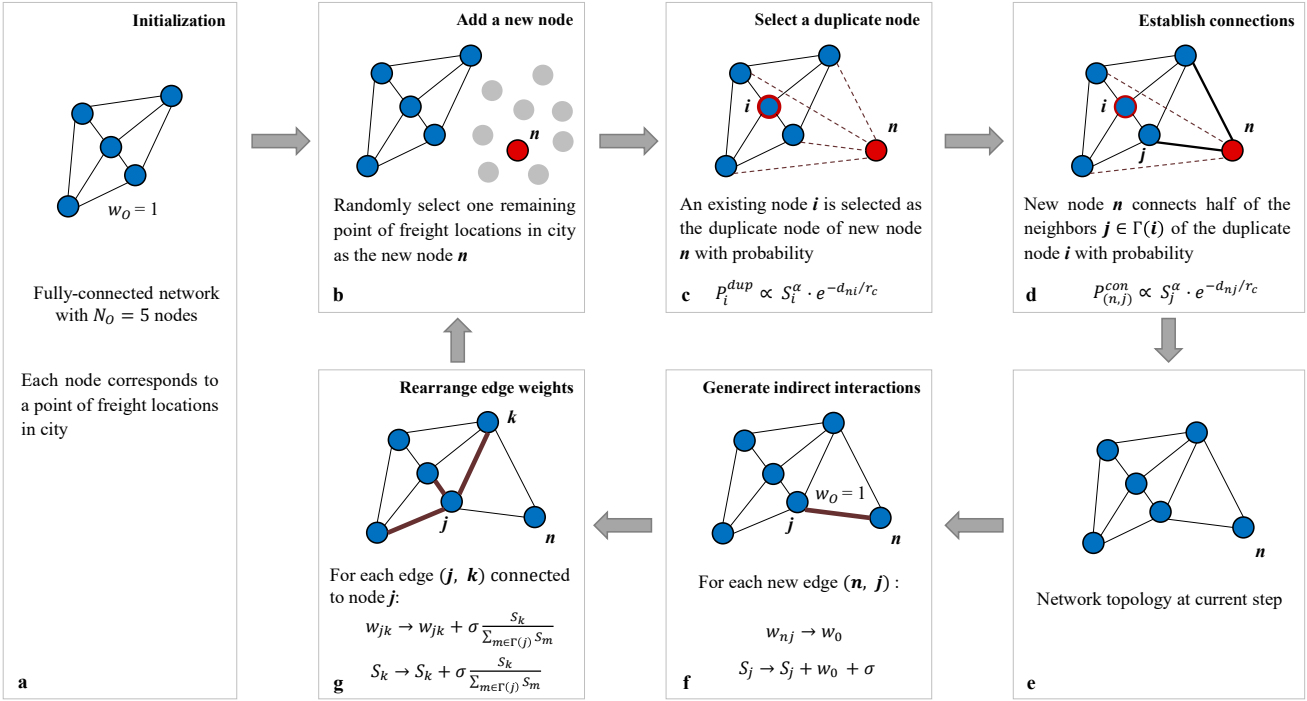

**Fig. 5**. Model illustration. The evolution of network at each step consists of two processes, i.e., topology growth (panel **b-d**) and weight updates (panel **e-g**). In the topology growth, a new node $n$ is first added (panel **b**) and selects its duplicate node by the duplication rule according to the gravity interaction pattern (panel **c**). The neighbors of duplicate node $i$ are the potential connecting nodes of the new node $n$. Next, the new node $n$ connects to half of the neighbors $\Gamma(i)$ of duplicate node $i$ by the connection rule according to the gravity interaction pattern, and this connection rule is formally similar to the preference attachment with spatial constraint (panel **d**). In weight update process starting from the current topology (panel **e**), network weights among nodes and edges are updated by the traffic increment rule, i.e., the establishment of a new edge of weight $w_0$ with the node $j$ generate a total indirect interaction $\sigma$ between new node $n$ and the neighbors $\Gamma(j)$ of node $j$ (panel **f**), and $\sigma$ is proportionally distributed among the edges departing from the node $j$ according to the strengths of neighbors $\Gamma(j)$ (panel **g**).

**Evolving network model**

We have quantitatively characterized a variety of structure properties of UHTMN in the empirical study above. However, how these structure properties are formed during the evolution of UHTMN is an important question that remains to be answered. To this end, we develop a weighted spatial network evolving model to explain the underlying dynamical mechanisms of UHTMN. The spatial interactions between freight locations are the core of the evolution of UHTMN. The spatial interaction pattern of gravity model[53] is the most widely used to analyze and predict the interactions between locations by integrating metrics of relative distance and size. In our model, we consider a duplication-based extension of gravity interaction pattern to explain the spatial interactions between freight locations during network evolution.

Our model starts with an initial fully connected seed network containing $N_0$ nodes (see **Fig. 5a**), which are randomly selected from among the freight locations in city. Each edge in this seed network is given a weight $w_0$. For simplicity, we set $N_0 = 5$ and $w_0 = 1$ both to constants. At each step, we randomly select a point from the remaining urban freight locations as the new added node $n$ (see **Fig. 5b**). Next, the new node $n$ selects interacting nodes from the current network and determines the corresponding interaction strengths. This process is equivalent to the emergence and development of a new freight location, e.g., company. The establishment of a new company is often accompanied by the duplication of the business modes of existing successful companies, i.e., the market operations of this new company are similar to those of existing successful companies[49,50]. The more similar the market operations of two companies are, the more they tend to transport cargo or provide products to the same clients. Therefore, our model considers system element duplication[46] in the topology growth, assuming that a node $i$ in the current network is selected as a duplicate node when the new node $n$ is added, i.e., all neighbors of this duplicate node $i$ are considered as potential interacting nodes of new node $i$ (see **Fig. 5c**). Then new node $i$ connects to some of these potential interacting nodes to establish edges (see **Fig. 5d**). In terms of selecting duplicate node, we consider the gravity interaction pattern between locations and define the duplication rule that an existing node $i$ is selected as the duplicate node of new node $n$ with probability

$$P_i \propto s_i^\alpha \cdot e^{-d_{ni}/r_c}, \quad (1)$$

where $s_i$ is the strength of node $i$, $\alpha$ is the attractiveness parameter, $d_{ni}$ is the Euclidean distance between nodes $n$ and $i$, and $r_c$ is the typical scale. This duplication rule imply that a new company tends to duplicate the business modes of spatially adjacent and attractive (or successful) companies, following the gravity interaction pattern. On the one hand, we measure the attractiveness of company by the variable $s_i^\alpha$, in which node strength $s_i$, i.e., freight transport volume is a main indicator of company attractiveness[54], and the exponent $\alpha$ embodies the effects of other indicators, such as profitability, firm assets and shipping prices, on the attractiveness of company. On the other hand, we measure the spatial proximity of two companies by the variable $e^{-d_{ni}/r_c}$, in which the typical scale $r_c$ controls the effects of trade, logistics and transaction costs associated with spatial distance on the location interaction.

In terms of establishing edges, we assume that new node $n$ connects to some but not all the neighbors $\Gamma(i)$ of duplicate node $i$, considering that commonly a new company would not have as many cooperation partners as the existing successful company. For simplicity, we assume that new node $n$ connects to half of the neighbors $\Gamma(i)$ of duplicate node $i$ (see **Fig. 5d**). New node $n$ selects which neighbors $\Gamma(i)$ to establish connections with also according to the gravity interaction pattern. The model considers the connection rule that each node $j \in \Gamma(i)$ is connected by the new node $n$ with probability

$$P_{(n,j)} \propto s_j^\alpha \cdot e^{-d_{nj}/r_c}, \quad (2)$$

where $s_j$ is the strength of node $j$, $d_{nj}$ is the Euclidean distance between nodes $n$ and $j$, $\alpha$ and $r_c$ are identical to equation (1). Variable $s_j^\alpha$ measures location attractiveness and variable $e^{-d_{nj}/r_c}$ measures the spatial proximity between locations. This connection rule is formally similar to the preference attachment with spatial constraint[42]. After establishing new edges, we can obtain the topology of the current network at this step (see **Fig. 5e**).

Next, we consider the weight updates after each new edge has been established. In the real world, when a new company establishes interaction with an existing company, it may also establish indirect interactions with other existing companies through this company, i.e. the traffic increment phenomenon[41]. For example, a logistics company could transport more goods to other companies in city through a new established connection with a freight hub, leading to the indirect interactions between this logistics company and other companies. Therefore, our model considers a traffic increment rule[41] for weight updates. We assume that the direct interaction strength of new node $n$ with connected node $j$ is $w_0 = 1$ (equal to the edge weights of initial network) and the indirect interaction strength of new node $n$ with other nodes via connected node $j$ is $\sigma$. The strength $s_j$ of the connected node $j$ is updated according to the rule (see **Fig. 5f**)

$$s_j \to s_j + w_0 + \sigma. \quad (3)$$

The indirect interaction strength $\sigma$ is assigned to the edges between the new node $n$ and the other nodes. Here we assume that new node $n$ only interacts indirectly with all neighbors $\Gamma(j)$ of the connected node $j$, and the indirect interaction strength of the new node $n$ with each node $k \in \Gamma(j)$ is proportional to the strength of node $k$. The weight $w_{jk}$ of each existing edge $(j, k)$ is rearranged according to the rule (see **Fig. 5g**)

$$w_{jk} \to w_{jk} + \sigma \frac{s_k}{\sum_{m \in \Gamma(j)} s_m}. \quad (4)$$

The strength $s_k$ of node $k \in \Gamma(j)$ is updated according to the rule (see **Fig. 5g**)

$$s_k \to s_k + \sigma \frac{s_k}{\sum_{m \in \Gamma(j)} s_m}. \quad (5)$$

In the next step, another new node is added from the remaining urban freight-related POIs, and the topology and weights of current network are evolved according to the above process. As the above evolving network model considers a duplication rule and a connection rule that are both based on the gravity interaction pattern, we refer to this model as the gravity duplication-connection model, i.e., GDC model.

**Table 2.** Global characterization of model networks for four municipalities directly under the central government in China. Estimated parameter $\alpha$, $r_c$ and $\sigma$; Number of nodes $N$; number of edges $E$; average degree $\langle k \rangle$; diameter $\phi$; average shortest path $\langle \ell \rangle$; average clustering coefficient $\langle C \rangle$; average shortest path $\langle \ell_{rand} \rangle$ and average clustering coefficient $\langle C_{rand} \rangle$ of ER random network that has the same number of nodes and edges as the model networks.

| City | $\alpha$ | $r_c$ | $\sigma$ | $N'$ | $E'$ | $\langle k \rangle'$ | $\phi'$ | $\langle \ell \rangle'$ | $\langle \ell_{rand} \rangle'$ | $\langle C \rangle'$ | $\langle C_{rand} \rangle'$ |
|---|---|---|---|---|---|---|---|---|---|---|---|
| Beijing | 0.057 | 11 | 1.7 | 15983 | 212594 | 26.6 | 6 | 3.92 | 3.27 | 0.0279 | 0.0017 |
| Shanghai | 0.116 | 14 | 1.9 | 16094 | 445832 | 55.4 | 8 | 3.01 | 2.77 | 0.0677 | 0.0034 |
| Chongqing | 0.072 | 22 | 1.7 | 10629 | 246989 | 46.5 | 9 | 3.32 | 2.76 | 0.1339 | 0.0044 |
| Tianjin | 0.144 | 21 | 2.0 | 10697 | 279274 | 52.2 | 7 | 2.87 | 2.70 | 0.1428 | 0.0049 |

**Model results**

We first estimate the optimal parameters, i.e., $\alpha$, $r_c$ and $\sigma$, of the GDC model by using the real data of each city (see **Methods** for details), and then compare the topological and weighted structure properties of the empirical UHTMN of a city with those of model network generated by the GDC model with the estimated optimal parameters. **Figure 2-4** show the model results for four municipalities directly under the central government in China, and those for other cities are shown in **Supplementary Note 3**.

With respect to topological structure properties, we compare the global topological properties, degree distribution $p(k)$, edge distance distribution $p(d)$, rich-club coefficient distribution $RC(k)$ and node neighbor similarity property of model networks with those of empirical UHTMN. We first calculate the global measures of the model networks generated by the GDC model for four Chinese municipalities directly under the central government in China, as shown in **Table 2**. For model networks, we obtain the average shortest path length $\langle \ell \rangle'$ close to that $\langle \ell_{rand} \rangle'$ of ER random graphs, which have the same number of nodes and edges as the model networks; the average clustering coefficient $\langle C \rangle'$ significantly larger than that $\langle C_{rand} \rangle'$ of the ER random graph. The results suggest that the GDC model can reproduce the small-world property of UHTMN. We then obtain the degree distributions $p(k)$ of model networks (see **Fig. 2a-d**), and they are in excellent agreement with the empirical results. This suggests the GDC model can reproduce the scale-free property of UHTMN, mainly attributed to the duplication rule and connection rule that are both based on the gravity interaction pattern, leading to a richer-get-richer topology growth in the model networks. Next, we obtain the edge distance distributions $p(d)$ of model networks (see **Fig. 2e-h**), and they are again in excellent agreement with the empirical results. This is mainly attributed to the gravity connection rule that makes a new node tends to establish connections with spatially adjacent nodes, reproducing the spatial feature of UHTMN. Moreover, the GDC model can reproduce the rich-club property of UHTMN (see **Fig. 2i-l**). This is also attributed to the gravity connection rule that makes nodes with large degrees tend to be well-connected to each other in the network growth. Further, the GDC model can reproduce the node neighbor similarity property of UHTMN (see **Fig. 2m-p**). The gravity duplication rule considered in the GDC model makes node pairs share many common neighbors as observed in the empirical UHTMN.

With respect to weighted structure properties, we compare the node strength distribution $p(s)$, edge weight distribution $p(w)$, weighted rich-club coefficient distribution $\rho^w(k)$, node average strength distribution $s(k)$, weighted and unweighted average degree of nearest neighbors $\langle k_{nn}\rangle^w(k)$ and $\langle k_{nn}\rangle(k)$, weighted and unweighted clustering coefficient $C^w(k)$ and $C(k)$ of model networks with those of empirical UHTMN. We first obtain the node strength distributions $p(s)$ (see **Fig. 3a-d**) and edge weight distributions $p(w)$ (see **Fig. 3e-h**) of model networks, and they are statistically indistinguishable from the empirical results. This is mainly attributed to the traffic increment rule considered in the GDC model, resulting in the heterogeneous distributions of network weights among nodes and edges in the model network. Moreover, the GDC model can reproduce the super-linear scaling relationship between the node average strength $s(k)$ and degree $k$, i.e., $s(k) \sim k^\beta$ ($\beta > 1$), in UHTMN (see **Fig. 3i-l**). This is also mainly attributed to the traffic increment rule that makes the node strengths grow faster than node degrees in the model network. In addition, the weighted rich-club coefficient distributions $\rho^w(k)$ are in good agreement with empirical results (see **Fig. 3m-p**). This is again mainly attributed to the traffic increment rule, which results in the significantly larger interaction strengths between well-connected large degree nodes. Furthermore, we calculate the weighted and unweighted average degree of nearest neighbors $\langle k_{nn}\rangle^w(k)$ and $\langle k_{nn}\rangle(k)$ (see **Fig. 4a-d**), weighted and unweighted clustering coefficient $C^w(k)$ and $C(k)$ (see **Fig. 4i-l**) of model networks. There is again an excellent agreement between these structural measures of model network and empirical UHTMN.

These results are again mainly attributed to the traffic increment rule, which makes the edges with higher weights tend to connect nodes with higher degrees that results in the distribution heterogeneity of network weights between nodes with different degrees.

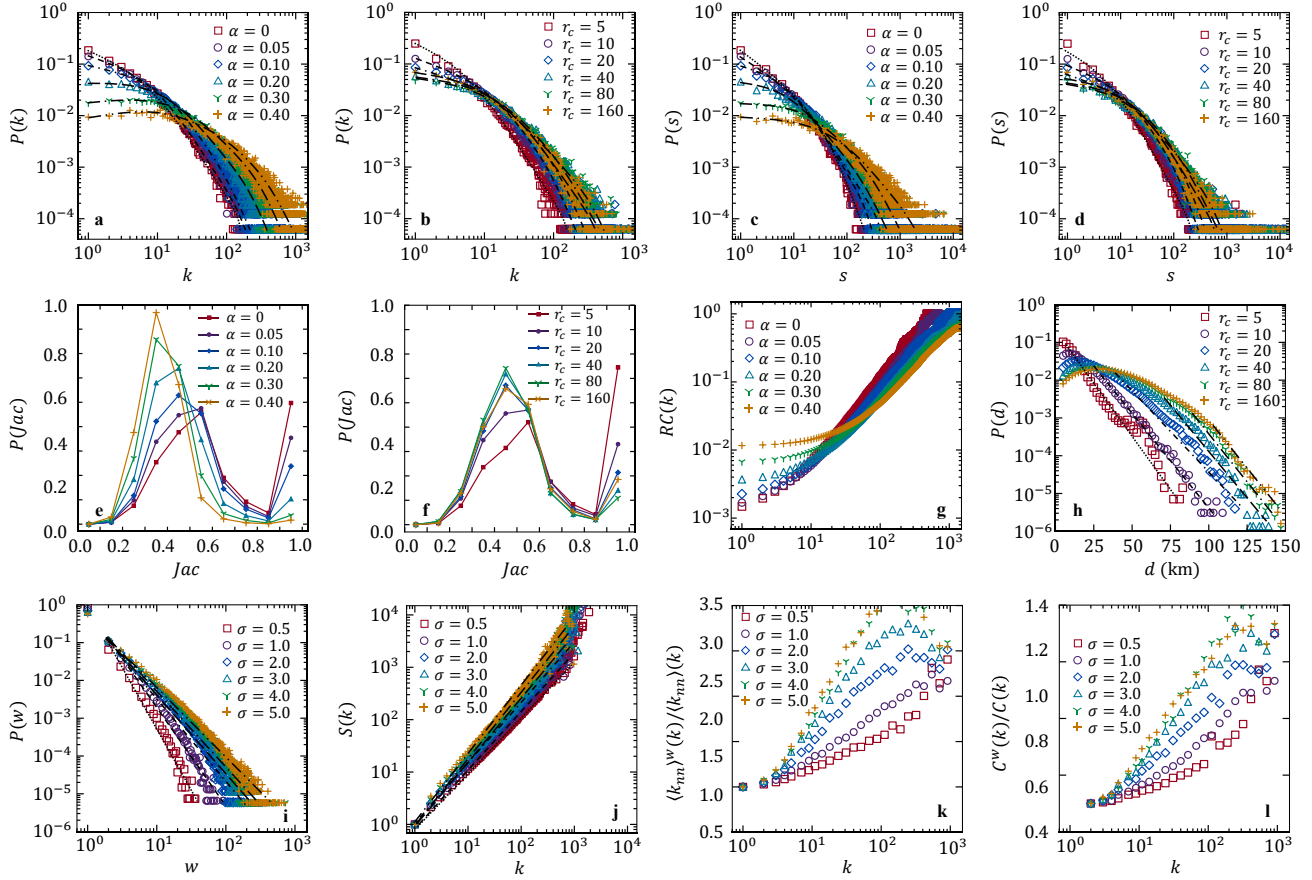

**Fig. 6**. Results of the model analysis. **a-b** Degree distributions $p(k)$ for different values of $\alpha$ and $r_c$. Each dashed line indicates a fitted cut-off power-law distribution in both panels. **c-d** Node strength distributions $p(s)$ for different values of $\alpha$ and $r_c$. Each dashed line indicates a fitted cut-off power-law distribution in both panels. **e-f** Distributions of Jaccard similarity coefficient of nodes with their similar nodes for different values of $\alpha$ and $r_c$. **g** Rich-club coefficient distributions $RC(k)$ for different values of $\alpha$. **h** Edge distance distribution $p(d)$ for different values of $r_c$. Each dashed line indicates a fitted exponential distribution. **i** Edge weight distributions $p(w)$ for different values of $\sigma$. Each dashed line indicates a fitted power-law distribution. **j** Average strength $s(k)$ as functions of degree $k$ for different values of $\sigma$. Each dashed line indicates a fitted super-linear relation. **K** The functions of $\langle k_{nn}\rangle^w(k)/\langle k_{nn}\rangle(k)$ with respect to degree $k$ for different values of $\sigma$. **l** The functions of $C^w(k)/C(k)$ with respect to degree $k$ for different values of $\sigma$.

## Model analysis

In the above, we have reproduced the various structure properties of empirical UHTMN by the GDC model with the estimated optimal parameters, and explained the underlying evolutionary mechanisms of UHTMN. Here, we further analyze the GDC model by exploring the effects of the variation of three parameters, i.e., $\alpha$, $r_c$ and $\sigma$, on the structure properties of model network. We take the empirical UHTMN in Beijing as the context for our analysis. For each analyzed parameter, we fix the other two parameters to be the optimal parameters ($\alpha = 0.057$, $r_c = 11$ and $\sigma = 1.7$) of the GDC model for Beijing UHTMN and set this analyzed parameter to different values ($\alpha = 0, 0.05, 0.10, 0.20, 0.30, 0.40$, $r_c = 5, 10, 20, 40, 80, 160$ and $\sigma = 0.5, 1.0, 2.0, 3.0, 4.0, 5.0$).

For each analyzed parameter, we generate 100 model networks (the number of nodes of them are equal to that of Beijing UHTMN) by the GDC model, and calculate the structure measures of these model networks on average. **Figure 6** shows the distributions of some main structure measures for different values of the analyzed parameters. The statistical results for all structure measures are presented in **Supplementary Note 4**. In the following, we analyze the effects of the model parameters on the network structure properties by these statistical results.

We first analyze the effects of the parameters on the scale-free property of the model network. We find that the degree distribution $p(k)$ is significantly affected by the parameter $\alpha$ (see **Fig. 6a**). As the parameter value $\alpha$ increases, a new node tends to duplicate and connect large degree nodes following the gravity interaction pattern, increasing the proportion of large degree nodes in model network, so the power exponent $\gamma$ decreases. Besides, the degree distribution $p(k)$ is also slightly affected by the parameter $r_c$ (see **Fig. 6b**). The smaller the parameter $r_c$, i.e., the stronger the effects of space, the more likely a new node is to establish short-distance connections and form local clusters. This increases the proportion of small degree nodes and thus makes the scale-free property more significant in the model network. The degree distribution $p(k)$ obeys a cut-off power-law distribution with respect to different values of $\alpha$ and $r_c$, and the power exponent $\gamma$ varies from 1 to 2. In the GDC model, there is a positive correlation between node strength and degree, so that the effects of model parameters on the strength distribution $p(s)$ are similar to those on the degree distribution $p(k)$ (see **Fig. 6c-d**).

Next, we analyze the effects of the parameters on the node neighbor similarity property, rich club property and spatial interaction distance of the model network. We find that the smaller the parameters $\alpha$ and $r_c$, the more significant the node neighbor similarity in the model network is (see **Fig. 6e-f**). This is because many spatially adjacent nodes with small degrees are well-connected at small $\alpha$ and $r_c$, raising the values of the Jaccard similarity coefficient between node pairs of the network. In addition, the larger the parameter $\alpha$, the higher the number of edges or network density. Therefore, an increase in parameter $\alpha$ will raise the rich-club coefficient $RC(k)$ with respect to small $k$ and slow down the growth of $RC(k)$ (see **Fig. 6g**), which means the rich-club property of network becomes insignificant. Further, the smaller the parameter $r_c$, the faster the edge distance distribution $p(d)$ decays exponentially (see **Fig. 6h**), which suggests the effects of space become stronger and make nodes more prone to establish short-distance interactions.

Finally, we analyze the effects of the parameters on the weighted structure properties of the model network. We find that the power-law decay behaviors of edge weight distribution $p(w)$ are mainly determined by the parameter $\sigma$, and the power exponent $\gamma_w$ is inversely correlated with the value of the parameter $\sigma$ (see **Fig. 6i**). Moreover, the larger the parameter $\sigma$, the larger the scaling exponent $\beta$ of the super-linear relationship between node strength $s(k)$ and degree $k$ (see **Fig. 6j**), implying a more significant traffic increment phenomenon in the model network. Further, as the parameter $\sigma$ increases, the functions of $\langle k_{nn}\rangle^w(k)/\langle k_{nn}\rangle(k)$ (see **Fig. 6k**) and $C^w(k)/C(k)$ (see **Fig. 6l**) with respect to degree $k$ grow faster. The above results suggest that the parameter $\sigma$ mainly determines the extent of distribution heterogeneity of network weights.

Taken together, the two parameters $\alpha$ and $r_c$, contained in both the duplication and connection rules following the gravity interaction pattern, mainly affect the topological properties of network; and the parameter $\sigma$ contained in the traffic increment rule mainly affect the weighted structure properties of network. The analysis of the effects of parameter variation on the network structure properties further shed light on the model mechanisms, and helps us understand the evolution of UHTMN more deeply.

**Discussion**

The past decade has witnessed extensive efforts into uncovering and understanding the structure properties and the evolutionary mechanisms of urban mobility networks[18,19,21,55,56]. However, in-depth studies on the structure and evolution of the UHTMN are still lacking. To address this research gap, we construct the UHTMNs of Chinese major cities by using massive urban heavy truck mobility data. We found that the UHTMNs of different Chinese cities have some prevalent structure properties, including scale-free, small-world, rich-club, distance distribution with exponential decay (spatial feature) and distribution heterogeneity of network weights, as also widely observed for the urban human mobility networks[14,25]. However, compared to other mobility networks (see **Supplementary Note 2** for details), UHTMNs have a distinctive structure property, i.e., the node neighbor similarity, implying many freight locations tend to transport cargo or provide products to the same clients. This phenomenon is due to the business mode similarity[49,50] between freight locations, by virtue of companies that drive a large share of movements of heavy trucks and thus shape the urban freight system. To the best of our knowledge, previous spatial network models[8] are not able to well explain the mechanisms underlying this phenomenon.

To explain the evolutionary mechanisms behind the wide range of structure properties of UHTMN, we develop a spatial network evolving model named the GDC model, which considers not only weight and space, but also system element duplication[46]. Under the duplication rule in the network growth, a new node first selects a duplicate node from among existing nodes and then establishes connections with some neighbors of this duplicate node. The new node and its duplicate node share many common neighbors, portraying the business mode similarity between freight locations. Previous spatial network models[8] considered the growth rule that a new node directly selects connecting nodes from all existing nodes, and thus were not able to reproduce the node neighbor similarity of a network. In addition, we consider the spatial interaction pattern between locations expressed by the gravity models[53] when the new node selects its duplicate node and connecting nodes. The gravity interaction patterns, i.e., spatial interactions between locations, are the results of a balance between location attractiveness and cost associated with space, as also considered in previous spatial network models. The GDC model can reproduce not only the node neighbor similarity property of UHTMN, but other prevalent structure properties that exist in mobility networks as well. Further, in the weight updates, we consider the traffic increment rule, which was first proposed in the BBV model[41]. This rule explains the real-world phenomenon that when a new company establishes interaction with an existing company, it may also establish indirect interactions with other existing companies through this company. Under this rule, the GDC model reproduces the super-linear relationship between node strength and degree, and the heterogeneity distribution of weights in UHTMN. Taken together, the GDC model considers the system element duplication in addition to weight and space compared to previous spatial network models, and thus is able to reproduce a variety of structure properties of UHTMNs of different cities. This also suggests that the GDC model is universal and captures the essential dynamics of the evolution of UHTMN.

The GDC model has many potential applications in practice. For the cities where massive heavy truck mobility data are available, we can use real data to construct the UHTMNs and estimate the parameters of the GDC model. We can further compare the disparities of urban freight systems in these cities by analyzing the differences in the estimated model parameters. As we discussed in the model construction, the values of the model parameters $\alpha$ and $r_c$ reflect the roles of location attractiveness and space in the spatial interaction pattern between locations. The value of the model parameter $\sigma$ reflects the indirect interaction strength between locations. The variations in model parameters across cities reveal the distinct characteristics of urban freight systems. This helps us to understand the current situation of economic development in specific cities and provide guidance for freight policy development. For cities where massive heavy truck mobility data are not available, we can approximate the model parameters by referring to the empirical UHTMNs of comparable cities in terms of land use, city size, economic development level, etc. The GDC model can be applied to simulate the UHTMNs of these cities with the estimated model parameters. Following on from this, the structure properties of the simulated UHTMN of specific cities help us understand the global characteristics of urban freight system, providing empirical support for the evaluation of freight management and planning measures.

The GDC model also provides new perspectives for modeling the evolution of many other real-world networks with the node neighbor similarity and spatial features. In addition to the UHTMN, the node neighbor similarity is also widely observed in other real-world networks, such as protein interaction networks, citation networks, business relationship networks and air transportation networks (see in **Supplementary Note 2**). Node pairs have many common neighbors implying that they share partners or have similar system functions. For example, two papers cite many of the same studies, possibly indicating that they belong to a related research field; two people serve on the same board of directors of multiple companies, meaning they may be business partners; and two genes determine the structure of several different proteins, implying the similarity of their biological functions. The node neighbor similarity properties observed in these real-world networks imply that the system element duplication may exist in the network's evolution, and this can be explained by the GDC model. Moreover, these real-world networks also have spatial features, implying that the interactions between nodes are constrained by the cost associated with space. For example, two spatially adjacent neurons in a neural network have a higher probability of being connected to reduce material and energy consumption[57]; the citation flows in the citation networks have been found to decrease with distance and obey the gravity law[58]. The network spatial features are also considered in the GDC model. In summary, the GDC model can help us to understand the evolution of a wide range of natural and social systems characterized by the node neighbor similarity and spatial features, and provide supports for regulating these real-world systems.

## Methods

### Datasets and network construction

The datasets we used to construct the UHTMN consist of heavy truck GPS trajectory data and urban freight-related point-of-interest (POI) data. Heavy truck

GPS trajectory data were from the China Road Freight Supervision and Service Platform (https://www.gghypt.net/). This platform is used to record the real-time geographic locations of all heavy trucks with a load exceeding 12 tons in China and monitor their traffic violations. Our GPS trajectory data contains 41 billion trajectory records of 2.6 million heavy trucks in China from May 18, 2018 to May 31, 2018. The attributes of trajectory records include truck ID, timestamp, longitude, latitude, speed and direction angle. Urban freight-related POI data were crawled from Amap (https://lbs.amap.com/), which is a leading map application in China. POI data provide the names and geographical coordinates of urban freight locations, including companies, markets, warehouses, railway stations and port terminals.

To construct the UHTMN, we need to obtain the truck flows between urban freight locations from heavy truck GPS trajectories. We first use the urban heavy truck trip origin-destination identification algorithm[32] to identify the trip chains of each heavy truck from its GPS trajectory, and then construct the individual mobility network of each heavy truck. The individual mobility network is an undirected weighted network where a node represents an urban freight location, an edge represents the movement of a truck that exists between two freight locations, and the edge weight represents the truck flow, i.e., the number of movements of this truck. Next, we obtain the truck flows between urban freight locations from heavy truck collective movements, and thus construct the UHTMN (see **Fig. 1a**). Network schematics of UHTMNs (see **Supplementary Data**) for different Chinese cities are shown in **Fig. 1b-e** and **Fig. S3**.

**Jaccard similarity coefficient**

Jaccard similarity coefficient[52] $Jac_{ij}$ is a common metric for measuring the similarity of nodes $i$ and $j$, defined as

$$Jac_{ij} = \frac{|\Gamma(i) \cap \Gamma(j)|}{|\Gamma(i) \cup \Gamma(j)|}, \quad (6)$$

where $\Gamma(i)$ denotes the set of neighbors of node $i$. The numerator denotes the cardinality of the intersection of $\Gamma(i)$ and $\Gamma(j)$, i.e., the number of common neighbors of nodes $i$ and $j$. The denominator denotes the cardinality of the concatenation of $\Gamma(i)$ and $\Gamma(j)$, i.e., the number of all neighbors of nodes $i$ and $j$. If $Jac_{ij} = 1$, then the neighbors of nodes $i$ and $j$ are identical; on the contrary, if $Jac_{ij} = 0$, then nodes $i$ and $j$ have no common neighbors.

**Model parameter estimation**

The GDC model contains three key parameters, i.e., the attractiveness parameter $\alpha$, typical scale $r_c$ and indirect interaction strength $\sigma$. We estimate the model parameters by using a graph similarity-based method[59] to reproduce the structure properties of empirical UHTMN as best as possible. This method estimates the optimal parameters $\boldsymbol{\theta}^* = (\alpha^*, r_c^*, \alpha^*)$ by maximising the similarity between model network $G_{Model(\boldsymbol{\theta})}$ and real network $G_{real}$. The similarity of two networks is measured by the distance between network attribute vectors. The estimation of the optimal parameters can be expressed as

$$\boldsymbol{\theta}^* = \underset{\boldsymbol{\theta}}{\mathrm{argmin}} \; \mathrm{d}\left(f(G_{Model(\boldsymbol{\theta})}), f(G_{real})\right), \quad (7)$$

where $f(G) = (g_1, \dots g_i \dots, g_n)$ is the network attribute vector of $G$ and $g_i$ denotes the calculated $i$-th attribute value. Constructing the network attribute vector $f(G)$ and selecting the distance function $\mathrm{d}(X, Y)$ are two key issues in the optimal parameter estimation. First, we construct the network attribute vector $f(G)$ by using three typical structure measures, i.e., node degree, edge distance and edge weight. For each structure measure, we calculate the probability distribution of this measure in logarithmic bins, and the probability in each bin corresponds to an attribute value in $f(G)$. In the end, we can obtain attribute vectors $f(G_{Model(\boldsymbol{\theta})})$ and $f(G_{real})$ of the same length consisting of the three structure measures. Second, we use the Canberra distance function[60] $\mathrm{d}(X, Y)$ to measure the distance between vectors $X$ and $Y$, expressed as

$$\mathrm{d}(X, Y) = \sum_{i=1}^{n} \frac{|x_i - y_i|}{|x_i| + |y_i|} \quad (8)$$

where $x_i$ and $y_i$ are the $i$-th element in the vectors $X$ and $Y$. In the model parameter estimation, we first obtain multiple optional parameter values for each of the three parameters $\alpha$, $r_c$ and $\sigma$ at certain intervals, and then use a grid approach to obtain the optional parameter set $\boldsymbol{\Theta} = (\boldsymbol{\theta_1}, \boldsymbol{\theta_2}, \dots, \boldsymbol{\theta_m})$, where $\boldsymbol{\theta_i} = (\alpha^i, r_c^i, \alpha^i)$ is $i$-th sample of optional parameter set. For each sample $\boldsymbol{\theta_i}$, we obtain 100 model networks generated by the GDC model with the parameters in $\boldsymbol{\theta_i}$ and to construct attribute vector $f(G_{Model(\boldsymbol{\theta_i})})$. The sample $\boldsymbol{\theta_i}$ ($i = 1, 2, \dots, m$) corresponding to the minimum Canberra distance between $f(G_{Model(\boldsymbol{\theta_i})})$ and $f(G_{real})$ calculated by equation (8) are the optimal parameters $\boldsymbol{\theta}^* = (\alpha^*, r_c^*, \alpha^*)$ as expressed in equation (7).

## Data availability

The authors declare that the data supporting the findings of this study are available within the paper and its Supplementary Information file, or from the authors upon reasonable request.

## Acknowledgements

B. Jia was supported by the National Natural Science Foundation of China (NSFC) under grant nos. 72288101 and 71971015. X. Y. Yan was supported by NSFC under grant nos. 72271019 and 71822102. Z. Gao was supported by NSFC under grant no. 72288101. We are grateful to Dr Z. Z. Yang for providing us the heavy truck mobility data, and to Dr D. Y. Zhi and Dr D. D. Song for data processing.

## Author contributions

Y. Yang, B. Jia, X. Y. Yan designed the research; Y. Yang, E. Liu and X. Y. Yan performed the research; Y. Yang, B. Jia and Z. Gao contributed analytic tools; Y. Yang and B. Jia analyzed the data; and Y. Yang, X. Y. Yan, M. de. Bok and L. A. Tavasszy wrote the paper.

## Competing interests

The authors declare no competing financial interests.

## References

1. Rodrigue J-P. *The geography of transport systems*, 5th edn. Routledge (2020).
2. Li R, Dong L, Zhang J, Wang X, Wang W-X, Di Z & Stanley HE. Simple spatial scaling rules behind complex cities. *Nature Communications* **8**, (2017).
3. Wei L, Chen G, Sun WJ & Li GQ. Recognition of operating characteristics of heavy trucks based on the identification of GPS trajectory stay points. *Security and Communication Networks* **2021**, (2021).
4. Aljohani K. Integrating logistics facilities in Inner Melbourne to alleviate impacts of urban freight transport. In: *Australasian Transport Research Forum (ATRF), 38th* (2016).
5. Schwartz S & Fleming S. Motor carrier safety: A statistical approach will better identify commercial carriers that pose high crash risks than does the current federal approach. *US Government Accountability Office, Washington DC, USA Publication GAO-07-585*, (2007).
6. Brelsford C, Lobo J, Hand J & Bettencourt LMA. Heterogeneity and scale of sustainable development in cities. *Proceedings of the National Academy of Sciences of the United States of America* **114**, 8963-8968, (2017).
7. Tavasszy L & De Jong G. *Modelling freight transport*, 1st edn. Elsevier (2014).
8. Barthelemy M. Spatial networks. *Physics Reports* **499**, 1-101, (2011).
9. Louail T, Lenormand M, Picornell M, Cantu OG, Herranz R, Frias-Martinez E, Ramasco JJ & Barthelemy M. Uncovering the spatial structure of mobility networks. *Nature Communications* **6**, (2015).
10. Gadar L, Kosztyan ZT, Telcs A & Abonyi J. A multilayer and spatial description of the Erasmus mobility network. *Scientific Data* **7**, (2020).
11. Di Clemente R, Luengo-Oroz M, Travizano M, Xu SR, Vaitla B & Gonzalez MC. Sequences of purchases in credit card data reveal lifestyles in urban populations. *Nature Communications* **9**, (2018).
12. Chang S, Pierson E, Koh PW, Gerardin J, Redbird B, Grusky D & Leskovec J. Mobility network models of COVID-19 explain inequities and inform reopening. *Nature* **589**, 82-U54, (2021).
13. Barbosa H, Barthelemy M, Ghoshal G, James CR, Lenormand M, Louail T, Menezes R, Ramasco JJ, Simini F & Tomasini M. Human mobility: Models and applications. *Physics Reports* **734**, 1-74, (2018).
14. Valdano E, Okano JT, Colizza V, Mitonga HK & Blower S. Using mobile phone data to reveal risk flow networks underlying the HIV epidemic in Namibia. *Nature Communications* **12**, (2021).
15. Pappalardo L, Simini F, Rinzivillo S, Pedreschi D, Giannotti F & Barabasi A-L. Returners and explorers dichotomy in human mobility. *Nature Communications* **6**, (2015).
16. Bassolas A, Barbosa-Filho H, Dickinson B, Dotiwalla X, Eastham P, Gallotti R, Ghoshal G, Gipson B, Hazarie SA, Kautz H, Kucuktunc O, Lieber A, Sadilek A & Ramasco JJ. Hierarchical organization of urban mobility and its connection with city livability. *Nature Communications* **10**, (2019).
17. Lenormand M, Louail T, Cantu-Ros OG, Picornell M, Herranz R, Murillo Arias J, Barthelemy M, San Miguel M & Ramasco JJ. Influence of sociodemographics on human mobility. *Scientific Reports* **5**, (2015).
18. Simini F, Gonzalez MC, Maritan A & Barabasi A-L. A universal model for mobility and migration patterns. *Nature* **484**, 96-100, (2012).
19. Yan X-Y, Wang W-X, Gao Z-Y & Lai Y-C. Universal model of individual and population mobility on diverse spatial scales. *Nature Communications* **8**, (2017).
20. Song C, Koren T, Wang P & Barabasi A-L. Modelling the scaling properties of human mobility. *Nature Physics* **6**, 818-823, (2010).
21. Gonzalez MC, Hidalgo CA & Barabasi A-L. Understanding individual human mobility patterns. *Nature* **453**, 779-782, (2008).
22. Barabasi AL & Albert R. Emergence of scaling in random

networks. *Science* **286**, 509-512, (1999).
23. Watts DJ & Strogatz SH. Collective dynamics of 'small-world' networks. *Nature* **393**, 440-442, (1998).
24. Papadopoulos F, Kitsak M, Angeles Serrano M, Boguna M & Krioukov D. Popularity versus similarity in growing networks. *Nature* **489**, 537-540, (2012).
25. Arcaute E. Hierarchies defined through human mobility. *Nature* **587**, 372-373, (2020).
26. Kaluza P, Koelzsch A, Gastner MT & Blasius B. The complex network of global cargo ship movements. *Journal of the Royal Society Interface* **7**, 1093-1103, (2010).
27. Xu M, Pan Q, Muscoloni A, Xia H & Cannistraci CV. Modular gateway-ness connectivity and structural core organization in maritime network science. *Nature Communications* **11**, (2020).
28. Yang Y, Jia B, Yan X-Y, Li J, Yang Z & Gao Z. Identifying intercity freight trip ends of heavy trucks from GPS data. *Transportation Research Part E: Logistics and Transportation Review* **157**, 102590, (2022).
29. Zhou W-X, Wang L, Xie W-J & Yan W. Predicting highway freight transportation networks using radiation models. *Physical Review E* **102**, (2020).
30. Wang L, Ma J-C, Jiang Z-Q, Yan W & Zhou W-X. Gravity law in the Chinese highway freight transportation networks. *Epj Data Science* **8**, (2019).
31. Bombelli A, Santos BF & Tavasszy L. Analysis of the air cargo transport network using a complex network theory perspective. *Transportation Research Part E: Logistics and Transportation Review* **138**, (2020).
32. Yang Y, Jia B, Yan X-Y, Jiang R, Ji H & Gao Z. Identifying intracity freight trip ends from heavy truck GPS trajectories. *Transportation Research Part C: Emerging Technologies* **136**, (2022).
33. Dorogovtsev SN & Mendes JFF. Evolution of networks. *Advances in Physics* **51**, 1079-1187, (2002).
34. Cimini G, Squartini T, Saracco F, Garlaschelli D, Gabrielli A & Caldarelli G. The statistical physics of real-world networks. *Nature Reviews Physics* **1**, 58-71, (2019).
35. Colizza V, Flammini A, Serrano MA & Vespignani A. Detecting rich-club ordering in complex networks. *Nature Physics* **2**, 110-115, (2006).
36. Dorogovtsev SN, Mendes JFF & Samukhin AN. Size-dependent degree distribution of a scale-free growing network. *Physical Review E* **63**, (2001).
37. Zheng M, Garcia-Perez G, Boguna M & Serrano MA. Scaling up real networks by geometric branching growth. *Proceedings of the National Academy of Sciences of the United States of America* **118**, (2021).
38. Hao B, Jing Y, Zhang S, Zhou Y & Ieee Computer SOC. General BBV model of weighted complex networks. In: *International Conference on Communication Software and Networks*) (2009).
39. Pan Z, Li X & Wang X. Generalized local-world models for weighted networks. *Physical Review E* **73**, (2006).
40. Wang WX, Hu B, Wang BH & Yan G. Mutual attraction model for both assortative and disassortative weighted networks. *Physical Review E* **73**, (2006).
41. Barrat A, Barthelemy M & Vespignani A. Weighted evolving networks: Coupling topology and weight dynamics. *Physical Review Letters* **92**, (2004).
42. Barrat A, Barthelemy M & Vespignani A. The effects of spatial constraints on the evolution of weighted complex networks. *Journal of Statistical Mechanics: Theory and Experiment*, (2005).
43. Allard A, Serrano MA, Garcia-Perez G & Boguna M. The geometric nature of weights in real complex networks. *Nature Communications* **8**, (2017).
44. Popovic M, Stefancic H & Zlatic V. Geometric origin of scaling in large traffic networks. *Physical Review Letters* **109**, (2012).
45. Louf R, Jensen P & Barthelemy M. Emergence of hierarchy in cost-driven growth of spatial networks. *Proceedings of the National Academy of Sciences of the United States of America* **110**, 8824-8829, (2013).
46. Chung F, Lu LY, Dewey TG & Galas DJ. Duplication models for biological networks. *Journal of Computational Biology* **10**, 677-687, (2003).
47. Pandey PK, Singh M, Goyal P, Mukherjee A & Chakrabarti S. Analysis of reference and citation copying in evolving bibliographic networks. *Journal of Informetrics* **14**, (2020).
48. Evlampiev K & Isambert H. Conservation and topology of protein interaction networks under duplication-divergence evolution. *Proceedings of the National Academy of Sciences of the United States of America* **105**, 9863-9868, (2008).
49. Winter SG & Szulanski G. Replication as strategy. *Organization Science* **12**, 730-743, (2001).
50. Dunford R, Palmer I & Benveniste J. Business Model Replication for Early and Rapid Internationalisation The ING Direct Experience. *Long Range Planning* **43**, 655-674, (2010).
51. Erdős P & Rényi A. On the evolution of random graphs. *Publ Math Inst Hung Acad Sci* **5**, 17-60, (1960).
52. Murphy AH. The Finley affair: A signal event in the history of forecast verification. *Weather and Forecasting* **11**, 3-20, (1996).
53. Erlander S & Stewart NF. *The gravity model in transportation analysis: theory and extensions*. Vsp (1990).
54. Feo-Valero M, Garcia-Menendez L & Garrido-Hidalgo R. Valuing freighttransport time using transport demand modelling: A bibliographical review. *Transport Reviews* **31**, 625-651, (2011).
55. Moro E, Calacci D, Dong X & Pentland A. Mobility patterns are associated with experienced income segregation in large US cities. *Nature Communications* **12**, (2021).
56. Ren Y, Ercsey-Ravasz M, Wang P, Gonzalez MC &

Toroczkai Z. Predicting commuter flows in spatial networks using a radiation model based on temporal ranges. *Nature Communications* **5**, (2014).
57. Bullmore E & Sporns O. Complex brain networks: graph theoretical analysis of structural and functional systems. *Nature Reviews Neuroscience* **10**, 186-198, (2009).
58. Pan RK, Kaski K & Fortunato S. World citation and collaboration networks: uncovering the role of geography in science. *Scientific Reports* **2**, 902, (2012).
59. Sala A, Cao L, Wilson C, Zablit R, Zheng H & Zhao BY. Measurement-calibrated graph models for social network experiments. In: *Proceedings of the 19th international conference on World wide web*). Association for Computing Machinery (2010).
60. Lance GN & Williams WT. Computer programs for hierarchical polythetic classification ("similarity analyses"). *The Computer Journal* **9**, 60-64, (1966).